\documentclass[twocolumn,separability,Amsterdam,nobibnotes, aps, pra,showpacs,preprintnumbers]{revtex4-1}
\usepackage{amssymb}
\usepackage{array}
\usepackage{graphicx}

\def\be{\begin{equation}}
\def\ee{\end{equation}}
\def\bea{\begin{eqnarray}}
\def\eea{\end{eqnarray}}

\begin{document}

\title{Geometrothermodynamics as a singular conformal thermodynamic geometry }%

\author{Seyed Ali Hosseini Mansoori$^{1}$}
\email{shossein@ipm.ir \& shosseini@shahroodut.ac.ir}
\author{Behrouz Mirza$^{2}$}
\email{b.mirza@cc.iut.ac.ir}
\affiliation{$^1$Faculty of Physics, Shahrood University of Technology, P.O. Box 3619995161, Shahrood, Iran\\
$^{2}$Department of Physics,
Isfahan University of Technology, Isfahan 84156-83111, Iran}

\date{\today}%

\begin{abstract}
In this letter, we first redefine our formalism of the thermodynamic geometry introduced in \citep{reff12, reff13}  by changing coordinates of the thermodynamic space by means of Jacobian matrices.  We then show that the geometrothermodynamics (GTD) is conformally related to this new formalism of the thermodynamic geometry. This conformal transformation is singular at unphysical points were generated in GTD metric. Therefore, working with our  metric neatly excludes all unphysical points without imposing any constraints.

\end{abstract}


\maketitle

\section{Introduction}\label{a}

Hawking and Bekenstein were the first physicists to notice an analogy between black holes and common thermodynamic systems \cite{reff1,reff2}. In fact, by considering the surface gravity and the horizon area, respectively, as the temperature and the entropy, one can interpret the four laws of thermodynamics for a black hole system \cite{reff3}. However, the statistical origin of black hole thermodynamics is still a big challenging question.

Several attempts have been made in order to describe thermodynamic behaviors of a black hole by making use of  the Riemannian geometry \cite{reff4,reff5,reff6}. In particular, Weinhold's metric \cite{reff8} and Ruppeiner's metric \cite{reff9,reff10}, which are defined as the Hessian matrix of the internal energy and entropy, respectively;
were used to find a direct correspondence between curvature singularities and phase transitions.
 However, some contradictory examples \cite{reff5,reff11} lead to these metrics have received criticisms for not being Legendre invariant, but nevertheless we proposed a new formulation of the Ruppeiner's metric, by considering the thermodynamic potentials related to the mass (instead of the entropy) by Legendre transformations \cite{reff12,reff13}. We have proved that this new formalism of thermodynamic geometry represents a one-to-one correspondence between the divergences of the heat capacities and those of the curvature scalars.

On the other hand, Geometrothermodynamics (GTD) approach  was introduced as a Riemannian thermodynamic structure obeying Legendre invariance \cite{reff14}. However, in the special case of a phantom RN-AdS black hole, the GTD fails to explain the one-to-one correspondence between phase transitions and singularities of the scalar curvature \cite{reff15, reff16}. 
In fact, the indiscriminate use of the natural thermodynamic variables and modified variables in black hole thermodynamics, may lead to reveal some ambiguities in GTD \cite{reff17, reff18}. For instance, to overcome these inconsistencies found in Ref. \cite{reff15}, one needs to consider an extended thermodynamic space, in which the cosmological constant is assumed to be a thermodynamic variable, and impose that the corresponding fundamental equation is that of a homogeneous function defined on this extended thermodynamic space \cite{reff18}.
Moreover, extra singularities appeared in GTD method \cite{reff16} can be interpreted as physical phase transitions at the level of response functions and stability changes \cite{Quevedo:2016swn}.

The aim of this letter is to find the source of these  inconsisitencies (unphysical points) which were generated in GTD by performing a conformal transformation between our formalism of thermodynamic geometry \cite{reff12,reff13} and GTD method. In point of fact, the roots of this conformal factor give us unphysical points. In order to construct this conformal transformation, we need to redefine our formalism in the GTD language by changing coordinates of the thermodynamic manifold. We also prove that the singularity property of this conformal factor leads to disappear physically the unphysical points, which are generated in GTD method, from curvature divergences.

The outline of this letter is as follows. In Section \ref{sec2}, we revisit our previous works about thermodynamic geometry and try to build the new formalism thermodynamic geometry by changing coordinates. In Section \ref{sec3}, we obtain a  conformal transformation between our new formalism and  GTD metric. In Section \ref{sec3} we study the thermodynamic geometry of the phantom RN-AdS black hole, and compare results of our metric  with the GTD metric. Finally, Section \ref{sec4} is devoted to discussions of our results.

  \section{The New formalism of the Thermodynamic geometry  } \label{sec2}
 Let us first review our previous results \cite{reff12, reff13,reff23} and then try to rewrite our formalism of the thermodynamic geometry in a new form which is suitable for our purpose.  

  In \cite{reff12, reff13}, we have proved that there is a one-to-one correspondence between divergences of the heat capacities at a fixed electric charge, $C_{Q}\equiv T{{\left( \frac{\partial S}{\partial T} \right)}_{Q}}$, and those of the curvature scalar, $\bar{R}(S, \Phi)$, by defining the following metric,
 \begin{equation} \label{ee12}
 \overline{g}_{ij}=\frac{1}{T}\frac{\partial^2\bar{M}}{\partial X^i \partial X^j}
 \end{equation}
 where ${{X}^{i}}=(S,\Phi)$ and $\overline{M}(S,\Phi )=M(S,Q(S,\Phi))-\Phi Q(S,\Phi)$  is the Enthalpy potential associated with the mass potential $M(S,Q)$ by Legendre transformation.
 Notice that for two dimensional thermodynamic space, the denominator of $\overline{R}(S,\Phi )$ is proportional to the square of the metric determinant, i.e., $\overline{R} \propto (\overline{g})^{-2}$. According to the first law of thermodynamics for Enthalpy potential, $
{d\overline{M}}(S,\Phi )=T dS-Q d\Phi $, the metric matrix is given by,
\begin{equation}\label{eeeq12}
\overline{\textbf{g}}=\frac{1}{T}\left( \begin{array}{cc}
  \noalign{\medskip}  {{\left( \frac{\partial T}{\partial S} \right)}_{\Phi }} & {{\left( \frac{\partial T}{\partial \Phi} \right)}_{S}}  \\
    \noalign{\medskip} {{-\left( \frac{\partial Q}{\partial S} \right)}_{\Phi}} & {{-\left( \frac{\partial Q}{\partial \Phi} \right)}_{S}}  \\
 \end{array} \right)=\frac{\partial(T,-Q)}{T\partial{(S,\Phi)}}
\end{equation}
Thus its determinant reads
\begin{eqnarray} \label{h22}
\nonumber \overline{g}&=&det({\overline{\textbf{g}}})=det \left[\frac{\partial(T,-Q)}{T\partial{(S,\Phi)}} \right]=det \left[\frac{\partial(T,-Q)}{T\partial{(S,\Phi)}}  \frac{\partial(S,Q)}{\partial{(S,Q)}} \right] \\
&=&det \left[\frac{\partial(T,-Q)}{T\partial{(S,
Q)}} \right]det\left[ \frac{\partial(S,Q)}{\partial{(S,\Phi)}} \right]=-\frac{C_S}{T^2C_Q}.
\end{eqnarray}
where $C_{s} \equiv {{\left( \frac{{\partial Q}}{{\partial \Phi }} \right)}_{S}}$ \cite{reff13}. Moreover, we have used the identity matrix in the first line. 
 The equality (\ref{h22}) indicates that the phase transitions of ${{C}_{Q}} (S, Q(S,\Phi))$ occur exactly at the singularities of $\overline{R}(S,\Phi )$.
Nevertheless, in some black hole systems, writing $Q$ as a function of $(S, \Phi)$ is usually hard or impossible. Therefore, it is convenient to rewrite the metric (\ref{ee12}) from coordinates $(S, \Phi)$ in the coordinate $(S,Q)$. Thus allow us transfer from the coordinates $(S,\Phi)$ to $(S,Q)$ by using below Jacobian matrix.
 \begin{equation}\label{eq1}
 \textbf{J}\equiv \frac{\partial \left( S,\Phi  \right)}{\partial \left( S,Q \right)}
 \end{equation}
In new coordinates, the metric elements must be changed by
 \begin{equation}\label{aa1}
 \overline{g}_{ij}'=J_{ik}^{T} \,\ \overline{g}_{kl} \,\ {{J}_{lj}}
 \end{equation}
 where $J^{T}$ is the transpose of $J$. Under varying coordinates, the metric (\ref{eeeq12}) takes the following form.
 \begin{eqnarray}\label{eq3}
\nonumber {{\bar{{\textbf{g}}'}}}&=&{{\left( \frac{\partial \left( S,\Phi  \right)}{\partial \left( S,Q \right)} \right)}^{T}}\left( \frac{\partial \left( T,-Q \right)}{T\partial \left( S,\Phi  \right)} \right)\left( \frac{\partial \left( S,\Phi  \right)}{\partial \left( S,Q \right)} \right)\\
 &=& {{\left( \frac{\partial \left( S,\Phi  \right)}{\partial \left( S,Q \right)} \right)}^{T}}\left( \frac{\partial \left( T,-Q \right)}{T\partial \left( S,Q \right)} \right)
 \end{eqnarray}

  After using Maxwell relation, $\Big(\frac{\partial T}{\partial Q}\Big)|_{S}=\Big(\frac{\partial \Phi}{\partial S}\Big)|_{Q}$, one gets from Eq.(\ref{eq3}) to the below relation for the metric elements.
 \begin{eqnarray}\label{2DM}
\nonumber  {{\bar{{\textbf{g}}'}}}&=&\text{diag}\Big(\frac{1}{T}\Big(\frac{\partial T}{\partial S}\Big)_{Q},-\frac{1}{T}\Big(\frac{\partial \Phi}{\partial Q}\Big)_{S}\Big)= \text{diag}(C_{Q}^{-1},-\frac{C_{S}^{-1}}{T})\\
&=& \frac{1}{T} \text{diag}\Big(\frac{\partial^2 M}{\partial S^2},-\frac{\partial^2 M}{\partial Q^2}\Big)
 \end{eqnarray}
In the last term, we have applied the first law of thermodynamic for mass potential, i.e. $dM=TdS+\Phi dQ$. It is surprising that after changing coordinates, the metric (\ref{ee12}) converts to a new form which is defined by the mass potential $M(S,Q)$. Taking advantage of this formalism, we can exploit easily the conformal transformation between our metric and GTD metric. 

 In the same way as above, for general black holes with $(n+1)$ thermodynamic variables, $X^{i}=\left(S,\Phi_{1},...,\Phi_{n}\right)$, and Enthalpy potential, $\overline{M}=M-\sum_{i}^{n}\Phi_{i}Q_{i}$, we define the metric matrix \cite{reff12, reff13} as
 \begin{equation}\label{ew12}
 \overline{\textbf{g}}=\frac{\partial \left( T,-Q_{1} ,-Q_{2},...,Q_{n} \right)}{T\partial \left( S,\Phi_{1} ,\Phi_{2},...,\Phi_{n}\right)}
 \end{equation}
where the first law of thermodynamic $d\bar{M}=TdS-\sum_{i}Q_{i}d \Phi_{i}$ have been used. By considering the Jacobian matrix,
 \begin{equation}
 \textbf{J}\equiv \frac{\partial \left( S,\Phi_{1},\Phi_{2},...,\Phi_{n}  \right)}{\partial \left( S,Q_{1},Q_{2},...,Q_{n} \right)}
 \end{equation}
under the coordinate change $\{S,\Phi_1,...,\Phi_{n}\} \to \{S,Q_1,...,Q_{n}\}$,
  the metric (\ref{ew12}) takes the below matrix block diagonal form.
 \begin{eqnarray}\label{EEEq1}
 \bar{{{\textbf{g}}}'}&=& \left( \begin{array}{cc}
  \noalign{\medskip}  C_{Q_{1},Q_{2},...,Q_{n}}^{-1}  & 0 \\
    \noalign{\medskip} 0  &  (-\textbf{G})_{n \times n} \\
 \end{array} \right)_{(n+1) \times (n+1)}
 \\
 \nonumber &=&\left( \begin{array}{cc}
  \noalign{\medskip}  \frac{1}{T}\frac{\partial^2 M}{\partial S^2}  & 0 \\
    \noalign{\medskip} 0  &  (-\textbf{G})_{n \times n} \\
 \end{array} \right)_{(n+1) \times (n+1)}
 \end{eqnarray}
 where $\textbf{G}$ is a square matrix of order $n$ which is given by
 \begin{equation}\label{eww1}
 \textbf{G}=\frac{1}{T}\frac{\partial^2M}{\partial Y^i \partial Y^j} \,\ ; \,\ Y^{i}=(Q_{1},Q_{2},...,Q_{n})
 \end{equation}
Therefore, the metric determinant in the new coordinates reads
 \begin{equation}
 \bar{{{g}}'}= \frac{G}{{-T}C_{Q_{1},Q_{2},...,Q_{n}} }
 \end{equation}
where $G=det(\textbf{G})$. This relation indicates the correspondence between the singularities of the Ricci scalar $\bar{R}$ and the phase transition of $C_{Q_{1},Q_{2},...,Q_{n}}$.

Form Eq. (\ref{eww1}), one can formulate Eq. (\ref{EEEq1}) as follows.
\begin{equation}\label{Rum1}
dl_{M}^2=-\frac{1}{T}\left( \eta_i^{ j} \, \frac{\partial^2M}{\partial X^j \partial X^l} \, d X^i d X^l \right)
\end{equation}
where $\eta_i^{ j}={\rm diag} (-1,1,...,1)$ and $X^{i}=(S,Q_{1},...,Q_{n})$ are extensive thermodynamic variables. This expression for the thermodynamic geometry shows that the singularities of the curvature made by thermodynamic potential $M(S,Q_{1},...,Q_{n})$, correspond to the phase transitions of $C_{Q_{1},Q_{2},...,Q_{n}}$.

 Furthermore, we have demonstrated that the phase transitions of specific heat ${{C}_{\Phi_{1},\Phi_{2},...,\Phi_{n} }}=T{{\left( \frac{\partial S}{\partial T} \right)}_{\Phi_{1},\Phi_{2},...,\Phi_{n} }} $ occur exactly at the singularities of the curvature $R(S,Q_{1},Q_{2},...,Q_{n})$ associated with the Ruppeiner metric \cite{reff12,reff13},
\begin{equation}\label{ee13}
{\textbf{g}}^{R}_{ij}=\frac{1}{T}\frac{\partial^2M}{\partial X^i \partial X^j} \hspace{1cm} X^{i}=\left(S,Q_{1},...,Q_{n}\right).
\end{equation}
  By making use of the first law of thermodynamic, $dM=TdS+\Phi_{1}dQ_{1}+...+\Phi_{n} dQ_{n}$, one can write down the Ruppeiner metric by
\begin{equation}
{\textbf{g}^{R}}=\frac{\partial \left( T,\Phi_{1} ,\Phi_{2},...,\Phi_{n} \right)}{T\partial \left( S,Q_{1} ,Q_{2},...,Q_{n}\right)}
\end{equation}
Under the coordinate change from $\{S,Q_1,...,Q_{n}\}$ to $ \{S,\Phi_1,...,\Phi_{n}\}$ through the below Jacobian matrix,
\begin{equation}
\textbf{J}\equiv \frac{\partial \left( S,Q_{1},Q_{2},...,Q_{n}  \right)}{\partial \left( S,\Phi_{1},\Phi_{2},...,\Phi_{n} \right)}
\end{equation}
 the metric element in new coordinate can be
written as follows.
  \begin{eqnarray}\label{EEEq2}
  {{{\textbf{g}}}'}&=&\left( \begin{array}{cc}
  \noalign{\medskip}  C_{\Phi_{1},\Phi_{2},...,\Phi_{n}}^{-1}  & 0 \\
    \noalign{\medskip} 0  &  (-\bar{\textbf{G}})_{n \times n} \\
 \end{array} \right)_{(n+1) \times (n+1)}
 \\
 \nonumber &=&\left( \begin{array}{cc}
  \noalign{\medskip}  \frac{1}{T}\frac{\partial^2 \bar{M}}{\partial S^2}  & 0 \\
    \noalign{\medskip} 0  &  (-\bar{\textbf{G}})_{n \times n} \\
 \end{array} \right)_{(n+1) \times (n+1)}
  \end{eqnarray}
Note that we have used the Maxwell relations in the above relation \cite{reff12}. Moreover, $\bar{\textbf{G}}$ is a square matrix of order $n$, i.e.,
 \begin{equation}
 \bar{\textbf{G}}=\frac{1}{T}\frac{\partial^2\bar{M}}{\partial Y^i \partial Y^j} \,\ ; \,\ Y^{i}=(\Phi_{1},\Phi_{2},...,\Phi_{n})
 \end{equation}
  In analogy with the metric (\ref{Rum1}), we can write down Eq.(\ref{EEEq2}) by
  \begin{equation}\label{Rubarm1}
dl_{\bar{M}}^2=-\frac{1}{T}\left( \eta_i^{ j} \, \frac{\partial^2\bar{M}}{\partial X^j \partial X^l} \, d X^i d X^l \right)
\end{equation}
 where $X^{i}=(S,\Phi_{1},...,\Phi_{n})$.
Now let us present the metrics (\ref{Rum1}) and (\ref{Rubarm1}) in the unit form as
\begin{equation}\label{Ru1}
dl_{NTG}^2=-\frac{1}{T}\left( \eta_i^{ j} \, \frac{\partial^2\Xi}{\partial X^j \partial X^l} \, d X^i d X^l \right)
\end{equation}
where $\eta_i^{ j}={\rm diag} (-1,1,...,1)$ and $\Xi$ is the thermodynamic potential. Here we have used the abbreviation NTG (New Thermodynamic Geometry) for this metric. Clearly, when we use the ensemble associated with the mass of the black hole, $\Xi=M$, the curvature obtained from our metric formalism (\ref{Ru1}) diverges exactly at the phase transitions of $C_{Q_{1},...,Q_{n}}$. Moreover, if we consider the ensemble associated with the enthalpy, $\Xi=\bar{M}$, curvature singularities give us the phase transitions of $C_{\Phi_{1},...,\Phi_{n}}$. It means that the phase transition structure of black holes depends on the ensemble. In next sections, we first introduce Geometrothermodynamics (GTD) metric and then try to derive a conformal factor between our thermodynamic geometry formulation (\ref{Ru1})  and GTD method.

\section{Conformal Transformation between the new formalism of the thermodynamic geometry and GTD}\label{sec3}
\label{sec1}
Let us consider the thermodynamic phase space with $(2 n+1)$ dimensions, $\mathcal{T}$, with independent coordinates $Z^{A}=\{\Xi ,X^{i},I^{i}\}$, $i=1... n$, where $\Xi$  represents the thermodynamic potential, and $X^{i}$ and  $I^{i}$ are the extensive and  intensive thermodynamic variables, respectively. Now, one can select on ${\cal T}$ a
non-degenerate metric $G=G(Z^A)$,
and the Gibbs 1-form $
\Theta = d\Xi - \delta_{ij} I^i d X^j$, in which  $\delta _{ij}$ is the kornecher delta.
 Moreover, the
metric $G$ is Legendre invariant if its functional dependence on $Z^A$
does not change under a Legendre transformation,
\bea
&& \{\Xi ,{{X}^{i}},{{I}^{i}}\}\to \{\tilde{\Xi },{{\tilde{X}}^{i}},{{\tilde{I}}^{i}}\}\\
\nonumber && \Xi = \tilde{\Xi }-{{\delta }_{ij}}{{\tilde{X}}^{i}}{{\tilde{I}}^{j}}\
\eea
Indeed, the Legendre invariance guarantees that the geometric properties
of $G$ do not depend on the thermodynamic potentials \cite{reff14}.
Also, the Gibbs 1-form $\Theta$ is Legendre invariant
in the sense that according to a Legendre transformation it behaves like $\Theta \to \tilde{\Theta}=d \tilde{\Xi}-\tilde{I}_{i} d\tilde{X}^{i}$.
The equilibrium space, $\mathcal{E}$ is then a subspace of $\mathcal{T}$ by means of the pullback $\phi^{*}$ which is associated with the embedding map $\phi: \mathcal{E} \to \mathcal{T}$ with constraint $\phi^{*} (\Theta)=0$ i.e.    $d\Xi=I_{i} dX^{j}$ which shows $\Xi=\Xi(X^{i})$ and $
I_{i}=\frac{d\Xi}{dX_{i}}$. Under these conditions, the general metrics form can be defined as
\bea
G^{GTD}=\Theta^2+({{\xi }_{ij}}{{X}^{i}}{{I}^{j}})({{\eta }_{lm}}d{{X}^{l}}d{{I}^{m}})
\eea
where ${{\eta }_{ij}}=diag(-1,1,....,1)$ and $\xi_{ij}$ is a diagonal constant matrix \cite{reff20,reff21}. Moreover,
pullback induces metric on $\mathcal{E}$ as
\begin{equation}\label{gII}
(dl^I)^{2}= \left(\xi^{j}_{i} X^i \frac{\partial \Xi}{\partial X^j}\right) \left( \eta_j^{ l} \, \frac{\partial^2\Xi}{\partial X^l \partial X^m} \, d X^j d X^m \right)
\end{equation}
 where $\eta_i^{ j}={\rm diag} (-1,1,...,1)$ and $\xi^{j}_{i}=\xi_{il} \delta^{jl}$ \cite{reff20,reff21}. It is obvious that components of the above metric
can be calculated explicitly once the thermodynamic potential $\Xi(X^{i})$ is given.
By using Euler’s identity, the conformal term can be put proportional to the potential $\Xi$. When we consider the case where $\Xi$ is homogeneous in $X^{i}$ of order $\beta$, i.e. $\Xi(\lambda X^{i})=\lambda^{\beta} \Xi(X^{i})$, then Euler’s identity satisfies $\beta \Xi=X^{i} \frac{\partial \Xi}{\partial X^{i}}$. Moreover, for the generic case where $\Xi$ is a generalized homogeneous function, i.e. $\Xi(\lambda^{\alpha_{i}}X_{i})=\lambda^{\alpha_{\Xi}} \Xi (X^{i})$, the Euler’s identity reads $\alpha_{i} X^{i} \frac{\partial \Xi}{\partial X_{i}}=\alpha_{\Xi} \Xi$.
Therefore, one can always choose the components of the diagonal matrix $\xi_{ij}$ proportional to $\alpha_{\Xi}$ so that the conformal factor becomes proportional to $\Xi$ as follows \cite{reff18, reff22}.
\be
(dl^{II})^{2} = \Xi \left( \eta_i^{ j} \, \frac{\partial^2\Xi}{\partial X^j \partial X^l} \, d X^i d X^l \right)
\label{gdown}
\ee

A comparison between Eqs. (\ref{gII}, \ref{gdown}) and Eq. (\ref{Ru1}) clarifies the conformal transformation between GTD metrics and our metric (\ref{Ru1}).  More precisely, we can exploit a conformally equivalent thermodynamic metrics as
\begin{equation}\label{TGGT}
g^{I,II}_{ij}=\Omega^{I,II}(X^i)^2 g^{NTG}_{ij}
\end{equation}
where
\begin{eqnarray}\label{EEEQ3}
\Omega^I(X^i)^2&=&-T \Big(\xi^{j}_{i} X^i \frac{\partial \Xi}{\partial X^j}\Big)\\
\Omega^{II}(E^a)^2&=&-T \Xi \label{EEEQ33}
\end{eqnarray}
and $g^{NTG}_{ij}=-\frac{1}{T}\eta_i^{ k} \, \frac{\partial^2\Xi}{\partial X^k \partial X^j}$. One can also find the below relationship between
scalar curvatures (see appendix A) from those metrics by
 \begin{eqnarray}\label{conformal11}
\nonumber R^{I,II}& \propto  &(\Omega^{I,II})^{-2} R^{NTG}
\end{eqnarray}
It is worth emphasizing that the above equation implies that
the scalar curvature $R^{I,II}$ has some extra singularity points coming from $\Omega^{I,II}=0$  in comparison with the scalar curvature $R^{NTG}$. Therefore, it is necessary to impose some physical constraints to get rid of these singularities.
However, it turns out that the conformal transformation  (\ref{EEEQ3}) and (\ref{EEEQ33}) are singular conformal transformation with physically different properties.
In fact, because these
transformations are singular at the points of extra (unphysical) roots, i.e. $\Omega^{I,II}=0$, the our formalism of the thermodynamic geometry physically
exclude these extra (unphysical) roots form the singularities of the curvature. In the light
of the above discussion, it should be noted that the conformal transformations are not invertible  since the Jacobians of the transformation, $(\Omega^{I,II})^2$,
vanish at the
(unphysical) roots.

In next section, we consider a phantom RN-AdS black hole as an extraordinary example to illustrate the important distinctions between our formalism and GTD.

\section{The phantom RN-AdS black hole}
\label{sec4}
The mass of a phantom RN-AdS black hole \cite{reff15} is expressed as a function of the thermodynamic variables $(S,Q)$ as
 \begin{equation} \label{ss2}
  M=\frac{1}{2}{{(S/\pi )}^{3/2}}(\frac{\pi }{S}-\frac{\Lambda }{3}+\frac{\eta {{\pi }^{2}}{{Q}^{2}}}{{{S}^{2}}})\
 \end{equation}
 where, $\Lambda$ is the cosmological constant, and $\eta= \pm 1$. As $\eta=1$, we have RN-AdS black hole solutions, while Phantom RN-AdS black hole solutions are obtained by choice of
$\eta=-1$ \cite{reff15}.
Moreover, $S=\pi r_{+}^{2}$ is the Bekenstein-Hawking entropy. According to the first law of thermodynamics, $dM=TdS+\Phi dQ$, the Hawking temperature,  $T$,  the electric potential,  $\Phi$,  and the specific heat capacity, $C_{Q}$, are given by
 \begin{equation} \label{mm2}
 T={(\frac{\partial M}{\partial S})}_{Q}=\frac{-\pi S+\Lambda {{S}^{2}}+\eta {{\pi }^{2}}{{Q}^{2}}}{-4{{(\pi S)}^{3/2}}}\
 \end{equation}
 \begin{equation}
 \Phi ={(\frac{\partial M}{\partial Q})}_{S}=\frac{{{(S/\pi )}^{3/2}}\eta {{\pi }^{2}}Q}{{{S}^{2}}}\
 \end{equation}
 \begin{equation}  \label{ee6}
 {{C}_{Q}}=T{{(\frac{\partial S}{\partial T})}_{Q}}=\frac{-2S(-\pi S+\Lambda {{S}^{2}}+\eta {{\pi }^{2}}{{Q}^{2}})}{(-\pi S-\Lambda {{S}^{2}}+3\eta {{\pi }^{2}}{{Q}^{2}})}
\end{equation}
It is obvious that the heat capacity $C_{Q}$ diverges at the values of the entropy $S_{1}=-( \pi / 2 \Lambda) (1+\sqrt{1+12 \eta \Lambda Q^2})$ and $S_{2}=-( \pi / 2 \Lambda) (-1+\sqrt{1+12 \eta \Lambda Q^2})$. More precisely, in the phantom RN-AdS black hole case, the value of the $S_{1}$ is only positive. Thus this case possesses just one point of phase transition.

It is also straightforward to check that $M$ is not homogeneous in $(S,Q)$ because one can not find a real $\beta$ such that $M (\lambda S, \lambda Q) = \lambda^{\beta} M(S,Q)$. We now apply our metric (\ref{Ru1}) to find the phase transitions of $C_{Q}$ through thermodynamic geometry. As mentioned, one needs to put the thermodynamic potential, $\Xi=M(S,Q)$ in Eq. (\ref{Ru1}) to check the phase transitions of $C_{Q}$. We therefore have
  \begin{equation}
  (dl^{NTG})^{2}=-\frac{1}{T}\left(-\frac{\partial^2 M}{\partial S^2}dS^2+\frac{\partial^2 M}{\partial Q^2}dQ^2\right)\
  \end{equation}
 In addition, the denominator of $R^{TG}$ is obtained by
 \begin{eqnarray} \label{ee5}
D(R^{NTG}) &=&(-S\pi  + \Lambda {{S}^{2}}+{{\pi }^{2}}\eta {{Q}^{2}}) \times \\
\nonumber &&{{(-S\pi  - \Lambda {{S}^{2}}+3{{\pi }^{2}}\eta {{Q}^{2}})}^{2}}
 \end{eqnarray}
 \begin{figure}[tbp]
 \includegraphics[scale=0.5]{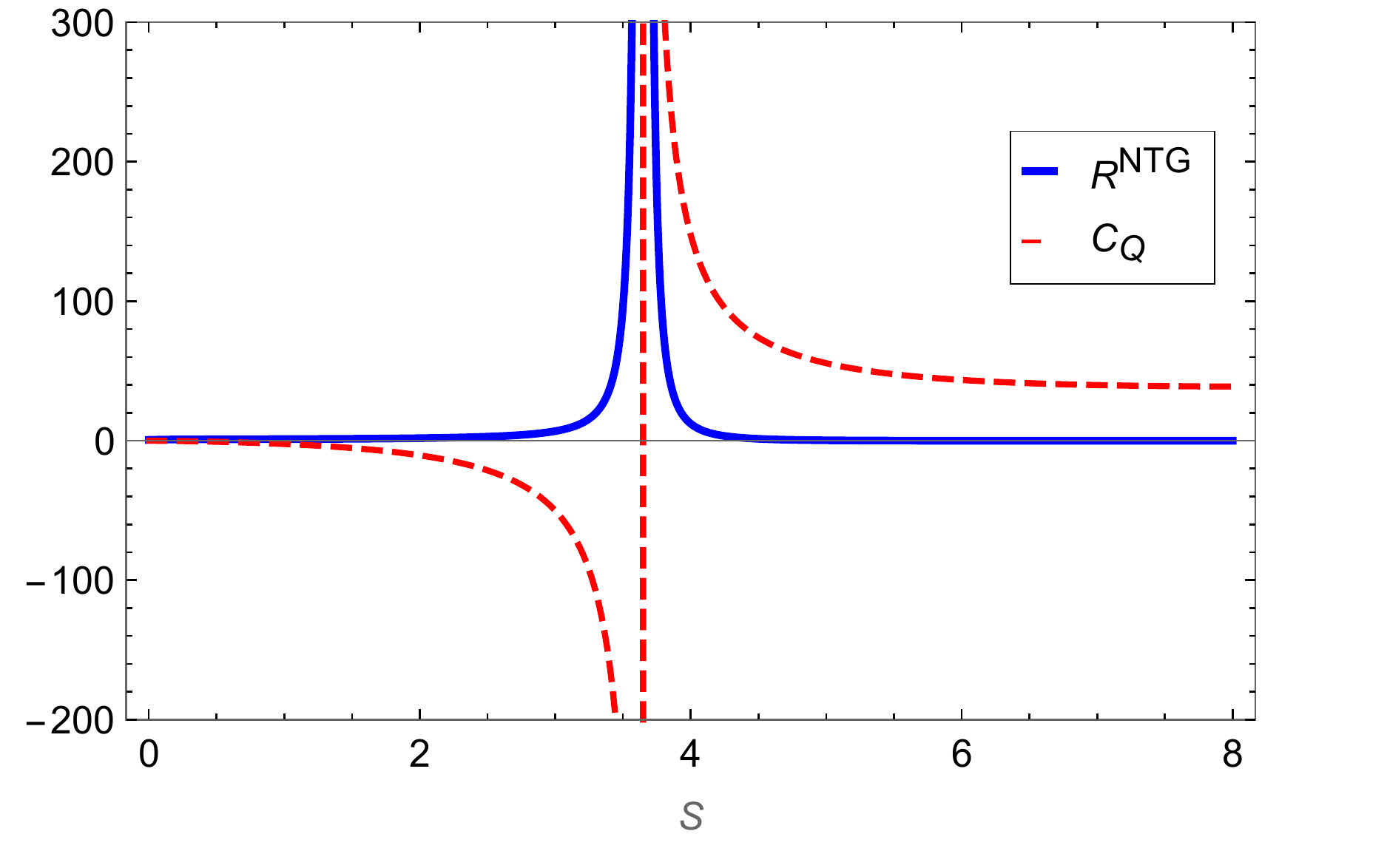}
 \caption{Diagram for the phase transition of $C_{Q}$ (dashed red curve) and singularities of the scalar curvature  $R^{NTG}(S,Q)$ (solid blue curve) versus entropy, $S$, for a phantom RN-AdS black hole with
  $Q=0.25$, $\Lambda=-1$, and $\eta=-1$. Note that roots of $T=0$ disappear in the phanton case.\label{fig1}   }
 \end{figure}
 \begin{figure}[tbp]
 \includegraphics[scale=0.49]{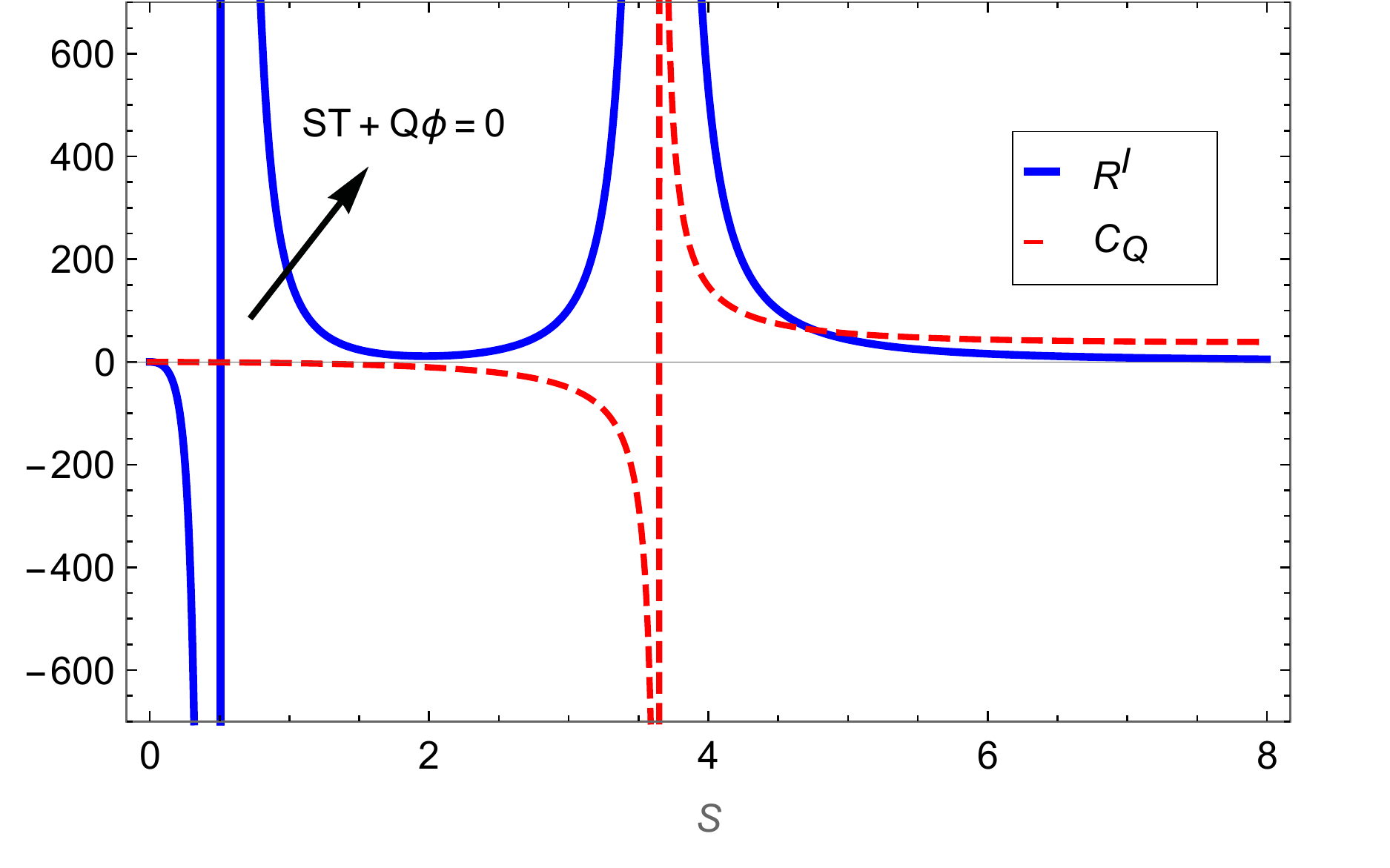}
 \caption{Diagram for the phase transition of $C_{Q}$ (dashed red curve) and singularities of the scalar curvature  $R^{I}(S,Q)$ (solid blue curve) versus entropy, $S$, for a phantom RN-AdS black hole with
  $Q=0.25$, $\Lambda=-1$, and $\eta=-1$.\label{fig2}   }
 \end{figure}
It is obvious that the first part of the denominator is zero only at the extremal limit ($T = 0$) which is forbidden by the third law of thermodynamics. Moreover, the roots of the second part give us all the phase transitions of $C_{Q}$. This point is also confirmed in Fig. (\ref{fig1}). As a consequence of our formalism for thermodynamic geometry, the curvature diverges exactly at the phase transitions with no other additional roots.

On the other hand, by substituting $\Xi=M$ and $X^{i}=(S,Q)$
in Eqs. (\ref{TGGT}) and (\ref{EEEQ3}), the line element of the GTD approach is given by
\begin{eqnarray}
(dl^{I})^{2}=-T\left(ST+Q\Phi\right) dl_{NTG}^2\label{me1}
\end{eqnarray}
Making use of Eq. (\ref{Eqqqqq5}), the denominator of the scalar curvature $R^{I}$ can be also calculated as
\begin{eqnarray}\label{GTDeq1}
D(R^{I})=\left(\Lambda S^2+\pi S-3\eta\pi^2 Q^2\right)^2 \\
\nonumber \times \left(\Lambda S^2-\pi S-3\eta\pi^2 Q^2\right)^3\,.\label{r1}
\end{eqnarray}
Clearly, the term in the leading parentheses of Eq. (\ref{GTDeq1}) presents phase transitions, whereas the term in the last parentheses describes the zero of the conformal factor, i.e. $ST+\Phi Q$. The same issue has been reperted  in \cite{REFF2} for RN black holes in presence of quintessence.
 As shown in Fig. (\ref{fig2}), it is obvious that the GTD approach gives us some extra singularity points which don't coincide with phase transition points. In other words, in the non-homogeneous potential case, Eq. (\ref{ss2}); the GTD metric is not able to provide a one-to-one correspondence between singularities and phase transitions. 

\begin{figure}[tbp]
 \includegraphics[scale=0.5]{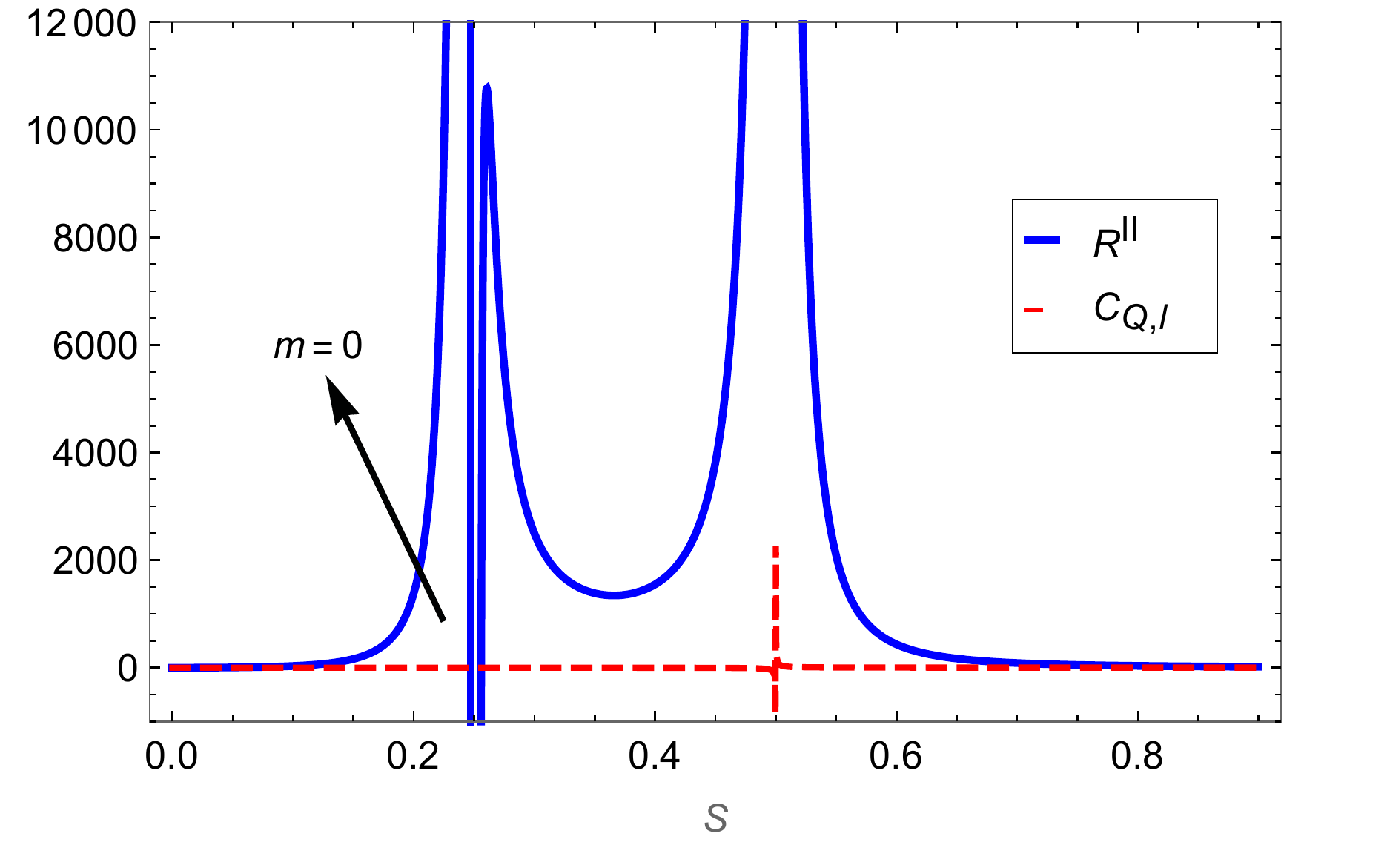}
 \caption{Diagram for the phase transition of $C_{Q,l}$ (dashed red curve) and the singularity of the scalar curvature  $R^{II}(s,Q,l)$ (solid blue curve) versus the new entropy variable, $s$, for a phantom RN-AdS black hole with
  $Q=0.25$, $l=1.73$, and $\eta=-1$. We have shown here the unphysical constraint $m=0$ on the plot which is ignored from the curveture singularies.\label{fig3}   }
 \end{figure}

In the remaining of this section, let us build a first-degree generalized homogeneous potential function from the fundamental mass (\ref{ss2}) and then give a brief explanation of the general characteristics of the GTD method and our metric (\ref{Ru1}), respectively. As discussed in \cite{reff18, reff19, reff17},  it is necessary to consider the cosmological constant as a thermodynamic variable \cite{REFF3} in order to make a homogeneous function. By rescaling
$S\rightarrow \lambda^{\alpha_S}S $, $Q\rightarrow \lambda^{\alpha_Q} Q$, and $\Lambda\rightarrow \lambda^{\alpha_\Lambda} \Lambda$ and assuming the conditions $\alpha_\Lambda = - \alpha_S$ and $\alpha_q = \frac{1}{2}\alpha_S$,
the fundamental mass (\ref{ss2}) is a generalized homogeneous function of degree $\beta= \alpha_S/2$.
For example, by replacing the cosmological constant $\Lambda$ by the AdS radius $l$ via $l^2= - 3/\Lambda$ and choosing $\alpha_S=1$, the mass formula will be a generalized homogeneous function of degree $1/2$, i.e.,
\be
M(\lambda S, \lambda^{1/2} l , \lambda^{1/2} Q) = \lambda^{1/2} M(S,l,Q)\ .
\ee
Note that one can reduce the degree of any generalized homogeneous function to one, by selecting the appropriate  variables \cite{REFF4}. By introducing the new entropy $s=\Big( {S}/{\pi}\Big)^{1/2}$, the mass (\ref{ss2}) converts to
\be
m = \frac{1}{2} s^3  \left(\frac{1}{s^2} +\frac{1}{l^2} +\eta \frac{Q^2}{s^4} \right)\ ,
\label{feq2}
\ee
It is easy to check that Eq. (\ref{feq2}) is a first-degree homogeneous function according to the Euler's identity. Now, form
the first law of thermodynamic, $dm=Tds+Ldl+\Phi dQ$; the heat capacity is given by
\be
C_{Q,l} = T\left(\frac{\partial S}{\partial T}\right)_{Q,l} =
\frac{s(3s^4+s^2 l^2 -\eta Q^2 l^2)}{2(3s^4 + \eta Q^2 l^2)} \ ,
\ee
which indicates the phase transition occurs at
$3s^4 =-\eta q^2 l^2$. By selecting the potential function $\Xi=m$, the GTD metric (\ref{gdown}) can be written as
\bea
\nonumber (dl^{II})^2&=& m \Big(-\frac{\partial^2 m}{\partial s^2}ds^2+\frac{\partial^2 m}{\partial Q^2}dQ^2+\frac{\partial^2 m}{\partial l^2}dl^2\\
&+&2 \frac{\partial^2 m}{\partial Q \partial l}dQ dl\Big) \ ,
\label{metfeq2}
\eea
it becomes manifest that, the equilibrium thermodynamic space must be extended to three dimensions when we consider the cosmological constant as a thermodynamic variable. Substituting Eq.(\ref{feq2}) in the above relation, the denominator of the scalar curvature is given by
\be\label{dq1}
D(R^{II}) = 3 (3s^4+\eta Q^2 l^2 )^2 (s^4 + s^2 l^2 +\eta Q^2 l^2)^3\
\ee
The term appearing in the first parenthesis in Eq. (\ref{dq1}) clarifies the phase transitions of $C_{Q,l}$, whereas the roots of the other parenthesis comes from $m=0$,
which is unphysical constraint, because in this case the first law of thermodynamics breaks down and the GTD metric can not defined at all \cite{reff18}. 
 Fig. (\ref{fig3}) illustrates the comparison between phase transitions of $C_{Q,l}$ and singularities of the $R^{II}$ for a phantom RN-AdS black hole. 

\begin{figure}[tbp]
 \includegraphics[scale=0.5]{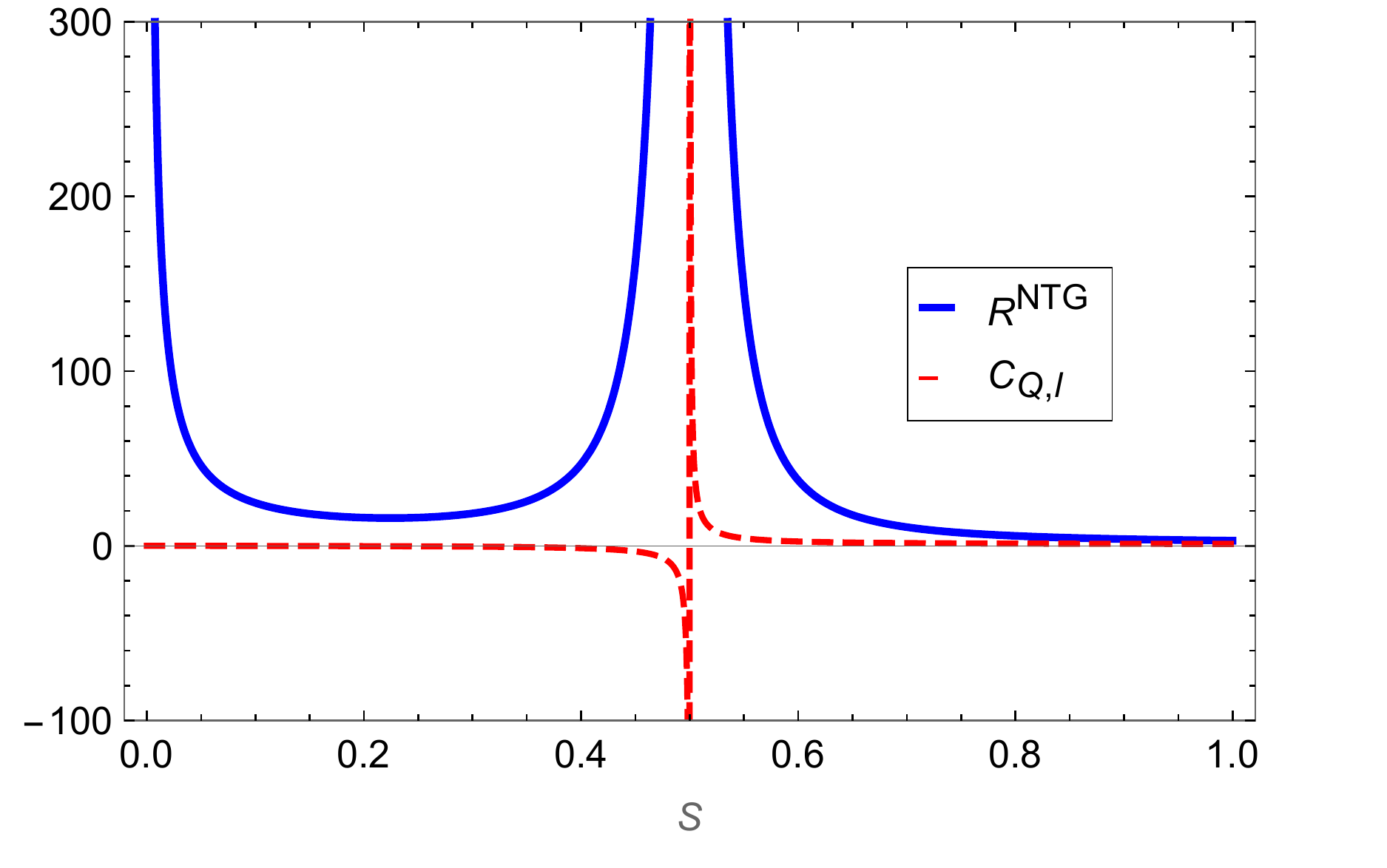}
 \caption{Diagram for the phase transition of $C_{Q,l}$ (dashed red curve) and the singularity of the scalar curvature  $R^{NTG}(s,Q,l)$ (solid blue curve) versus the new entropy variable, $s$, for a phantom RN-AdS black hole with
  $Q=0.25$, $l=1.73$, and $\eta=-1$.\label{fig4}   }
 \end{figure}

Now let us construct our metric for Phantom case. By considering $\Xi=m$ and $X^{i}=(s,l,Q)$ in our metric (\ref{Ru1}), we get
\bea
\nonumber dl_{NTG}^2&=& -\frac{1}{T} \Big(-\frac{\partial^2 m}{\partial s^2}ds^2+\frac{\partial^2 m}{\partial Q^2}dQ^2+\frac{\partial^2 m}{\partial l^2}dl^2\\
&+&2 \frac{\partial^2 m}{\partial Q \partial l}dQ dl\Big)
\eea
Therefore, the denominator of the scalar curvature reads
\be
D(R^{NTG}) = 4(3s^4+\eta Q^2 l^2 )^2.
\ee
It is interesting that the curvature singularity, $3s^4 =-\eta q^2 l^2$, give us just the phase transition point without imposing any constraints like $m=0$ (See Fig. (\ref{fig4})). In other words, our formalism physically excludes completely unphysical points $m=0$ which
appeared in the GTD metric. Therefore, compared to other techniques like GTD, our thermodynamic geometry formalism provides a powerful tool to achieve a one-to-one correspondence between singularities and phase transitions.
\section{conclusion}\label{sec5}
In this letter we rewrited our formalism of the thermodynamics geometry, previously introduced in \cite{reff12}, in the language of the general potential function $\Xi$ via Eq. (\ref{Ru1}). 
 It is worth mentioning that there is a one-to-one correspondence between the phase transition points of a black hole and singularities of the curvature associated with our metric (\ref{Ru1}).
  
  Moreover, the GTD metric is related to the new formalism of the thermodynamic geometry by means of a singular conformal transformation. 
   For a non-homogeneous potential function, we therefore proved that the roots of the conformal factor, $\xi^{j}_{i} X^i \frac{\partial \Xi}{\partial X^j}$, generate some singularity points which do not correspond to the phase transition points. On the other hand, in a homogeneous potential case, by imposing the physical constraint, $\Xi \neq0$,  the phase transitions occur exactly at curvature singularities of both metrics. 
   


In fact, unphysical points exclude physically form curvature singularities as one utilizes our formalism of the thermodynamic geometry rather than other thermodynamic geometric approaches like GTD method.

\section{Acknowledgments}

We are grateful to Mohamad Ali Gorji and Mustapha Azreg-A\"{\i}nou for extremely helpful discussions and comments about this work. We thank Mohamad Mahdi Davari Esfahani, Matteo Baggioli and Tsvetan Vetsov for reading a preliminary version of the draft. We appreciate the referee for his/her instructive comments.
\appendix
\section{Conformally equivalent metrics}\label{A1}
More generally, one can consider two conformally equivalent metrics as
\begin{equation}
\bar{g}_{\mu \nu}=\Omega^2 (x) g_{\mu \nu}
\end{equation}
where $\Omega (x)$ is called the conformal factor.
For the inverse metric and the metric determinant, we therefore have
\begin{equation}
\tilde{g}^{\mu \nu}=\Omega^{-2} g^{\mu \nu} \hspace{1cm} \tilde{g}=\Omega^{2D} g
\end{equation}
According to the new metric, the Ricci scalar can be written as \cite{reff24}
\begin{eqnarray}\label{conformal1}
\nonumber \tilde{R}&=&\Omega^{-2} \Big[R-2(D-1) \Box (Ln \Omega)\\
&-&(D-1) (D-2) \frac{g^{\alpha \beta} \nabla_{\alpha} \Omega \nabla_{\beta} \Omega}{\Omega^2}\Big]
\end{eqnarray}
for $D>2$, the latter formula can also be simplified by
\begin{equation}
\tilde{R}=\Omega^{-2}\Big[R-\frac{4 (D-1)}{D-2} \Omega^\frac{-(D-2)}{2} \Box \Omega^\frac{-(D-2)}{2}\Big]
\end{equation}
In the special case $D=2$, Eq.(\ref{conformal1}) leads to
\begin{equation}\label{Eqqqqq5}
\tilde{R}=\Omega^{-2} \Big[R-2 \Box (Ln \Omega)\Big]
\end{equation}




\end{document}